\documentclass[twocolumn,description,aps]{revtex4-1}
\usepackage{graphicx,epstopdf,amsmath,amssymb,multirow,CJK,color,tabularx}
\usepackage[german,english]{babel}
\usepackage[colorlinks, linkcolor=blue,anchorcolor=blue,citecolor=blue,urlcolor=blue]{hyperref}

\begin{document}

\title{\emph{R}RuB$_{2}$ (\emph{R}=Y, Lu), topological superconductor candidates with hourglass-type Dirac ring}

\author{Yan Gao$^{1}$}
\email{These two authors contributed equally to this work.}
\author{Peng-Jie Guo$^{1,2}$}
\email{These two authors contributed equally to this work.}
\author{Kai Liu$^{1}$}\email{kliu@ruc.edu.cn}
\author{Zhong-Yi Lu$^{1}$}\email{zlu@ruc.edu.cn}

\affiliation{$^{1}$Department of Physics and Beijing Key Laboratory of Opto-electronic Functional Materials $\&$ Micro-nano Devices, Renmin University of China, Beijing 100872, China}
\affiliation{$^{2}$Songshan Lake Materials Laboratory, Dongguan, Guangdong 523808, China}

\date{\today}

\begin{abstract}
Topological properties and topological superconductivity in real materials have attracted intensive experimental and theoretical attention recently. Based on symmetry analysis and first-principles electronic structure calculations, we predict that \emph{R}RuB$_{2}$ (\emph{R}=Y, Lu) are not only topological superconductor (TSC) candidates, but also own the hybrid hourglass-type Dirac ring which is protected by the nonsymmorphic space group symmetry. Due to the band inversion around the time-reversal invariant $\Gamma$ point in the Brillouin zone, \emph{R}RuB$_{2}$ also have Dirac topological surface states (TSSs). More importantly, their TSSs on the (010) surface are within the band gap of bulk and cross the Fermi level, which form single Fermi surfaces. Considering the fact that both YRuB$_{2}$ and LuRuB$_{2}$ are superconductors with respective superconducting transition temperatures ($\emph{T}_c$) of 7.6~K and 10.2~K, the superconducting bulks will likely induce superconductivity in the TSSs via the proximity effect. The ternary borides \emph{R}RuB$_{2}$ may thus provide a very promising platform for studying the properties of topological superconductivity and hourglass fermions in the future experiments.
\end{abstract}

\date{\today} \maketitle

\section{INTRODUCTION}\label{sec_introduction}

Topological superconductors (TSCs) with Majorana zero modes (MZMs) at boundaries and vortexes have attracted intense interest due to their exotic physical phenomena and potential applications in topological quantum computation~\cite{1SDSarma2008,2QiZhang,3YAndo}. To well explore the exotic properties of MZMs, it is necessary to search for various TSCs as many as possible, especially those with high superconducting transition temperature ($\emph{T}_c$) as well as high sample quality. However, intrinsic spinless $p$+i$p$ type TSCs are very scarce~\cite{4YMaeno}. Instead, a feasible scheme that can achieve the equivalent $p_x$+i$p_y$ type superconductivity has been proposed by Fu and Kane~\cite{5FuKane}, by which the topological superconductivity can be obtained in the topological surface states (TSSs) of a topological insulator (TI) via doping or proximity effect, such as Cu/Sr/Nb-doped Bi$_{2}$Se$_{3}$~\cite{6CavaHasan,7LiuZhang,8AsabaFuLi}, In-doped SnTe~\cite{9FuYAndo}, Bi$_{2}$Se$_{3}$/NbSe$_{2}$, and Bi$_{2}$Se$_{3}$/Bi2212 heterostructures~\cite{10JPXuDQian,11Schneeloch}.

In order to avoid the complex interface effect and lattice mismatch between a superconductor and a TI or the disorder introduced by doping, a natural idea is to realize topological superconductivity in single compounds~\cite{12Felser,13GuanChuang}. Although much effort has been devoted to searching for stoichiometric single-compound TSCs ~\cite{14JFZhang,15MKimKMHo,16KHJinFLiu}, for example, the existing superconductors Au$_{2}$Pb (1.18 K)~\cite{17Canfield}, PdTe$_{2}$ (1.7 K)~\cite{18NohPark}, PbTaSe$_{2}$ (3.8 K)~\cite{19PJChenHTJeng,20TRChang}, BiPd (3.8 K)~\cite{21Raychaudhuri,22SunWahl}, YD$_{3}$ (0.78~K to 4.72~K)~\cite{23XHTuBTWang}, $\beta$-PdBi$_{2}$ (5.3 K)~\cite{24SakanoIshizaka}, all the known TSC candidates are confronted with a very low $\emph{T}_c$. As the energy spacing of the Caroli-de Gennes-Matricon (CdGM) bound states~\cite{25CaroliMatricon} is in the order of $\frac{{\Delta}^{2}}{\varepsilon_f}$ (the $\Delta$ and $\varepsilon_f$ are the superconducting gap and the Fermi energy, respectively), increasing the $\emph{T}_c$ is beneficial to the experimental observation of the MZMs. An ideal stoichiometric single-compound TSC should meet three criteria at the same time: (1) The stoichiometric compound is an s-wave superconductor with relatively high $\emph{T}_c$; (2) The compound possesses topological surface states with spin-moment locking; (3) The TSSs remain separated from the bulk states and across the Fermi level. Unfortunately, the compounds that satisfy the criteria are still very rare so far.

In this work, we propose that the rare-earth transition-metal ternary borides \emph{R}RuB$_{2}$ (\emph{R}=Y, Lu) are ideal topological superconductor candidates that meet the above criteria.  Moreover, they also own hourglass-type nodal rings protected by the nonsymmorphic space group symmetries. Considering the fact that the experimental $\emph{T}_c$ of YRuB$_{2}$ and LuRuB$_{2}$ are 7.6~K and 10.2~K, respectively~\cite{26Kishimoto}, the ternary borides \emph{R}RuB$_{2}$ may be a promising platform for studying the exotic properties of the topological superconductivity and the hourglass fermions.

\section{METHOD}\label{sec_method}

The electronic structures of \emph{R}RuB$_{2}$ (\emph{R}=Y, Lu) were investigated based on the density functional theory (DFT) calculations using the projector augmented wave (PAW) method~\cite{27PAW} as implemented in the VASP package~\cite{28VASP}. The exchange-correlation functional was described by the generalized gradient approximation (GGA) of the Perdew-Burke-Ernzernof (PBE) type~\cite{29PBE}. The kinetic energy cutoff of the plane wave basis was set to 500~eV. An $8\times8\times8$ Monkhorst-Pack $k$-point mesh~\cite{30kmesh} was taken for the Brillouin zone (BZ) sampling, and the Gaussian smearing method with a width of 0.05~eV was adopted. The energy and force convergence criteria were set to $10^{-6}$~eV and 0.001~eV/\text{\AA}, respectively. The optimized lattice parameters for YRuB$_{2}$ are $a=5.914~\text{\AA}$, $b=5.320~\text{\AA}$, and $c=6.357~\text{\AA}$, and for LuRuB$_{2}$ are $a=5.818~\text{\AA}$, $b=5.252~\text{\AA}$, and $c=6.282~\text{\AA}$, in good agreement with their experimental values ($a=5.910~\text{\AA}$, $b=5.306~\text{\AA}$, and $c=6.363~\text{\AA}$ for YRuB$_{2}$ and $a=5.809~\text{\AA}$, $b=5.229~\text{\AA}$, and $c=6.284~\text{\AA}$ for LuRuB$_{2}$)~\cite{31RRuB2lattice}. The surface states and spin textures of a semi-infinite \emph{R}RuB$_{2}$ (\emph{R}=Y, Lu) surface were calculated with the surface Green's function method by using the WannierTools~\cite{32WannierTools} package.

\begin{figure}[t]
	\centering
	\includegraphics[width=0.48\textwidth]{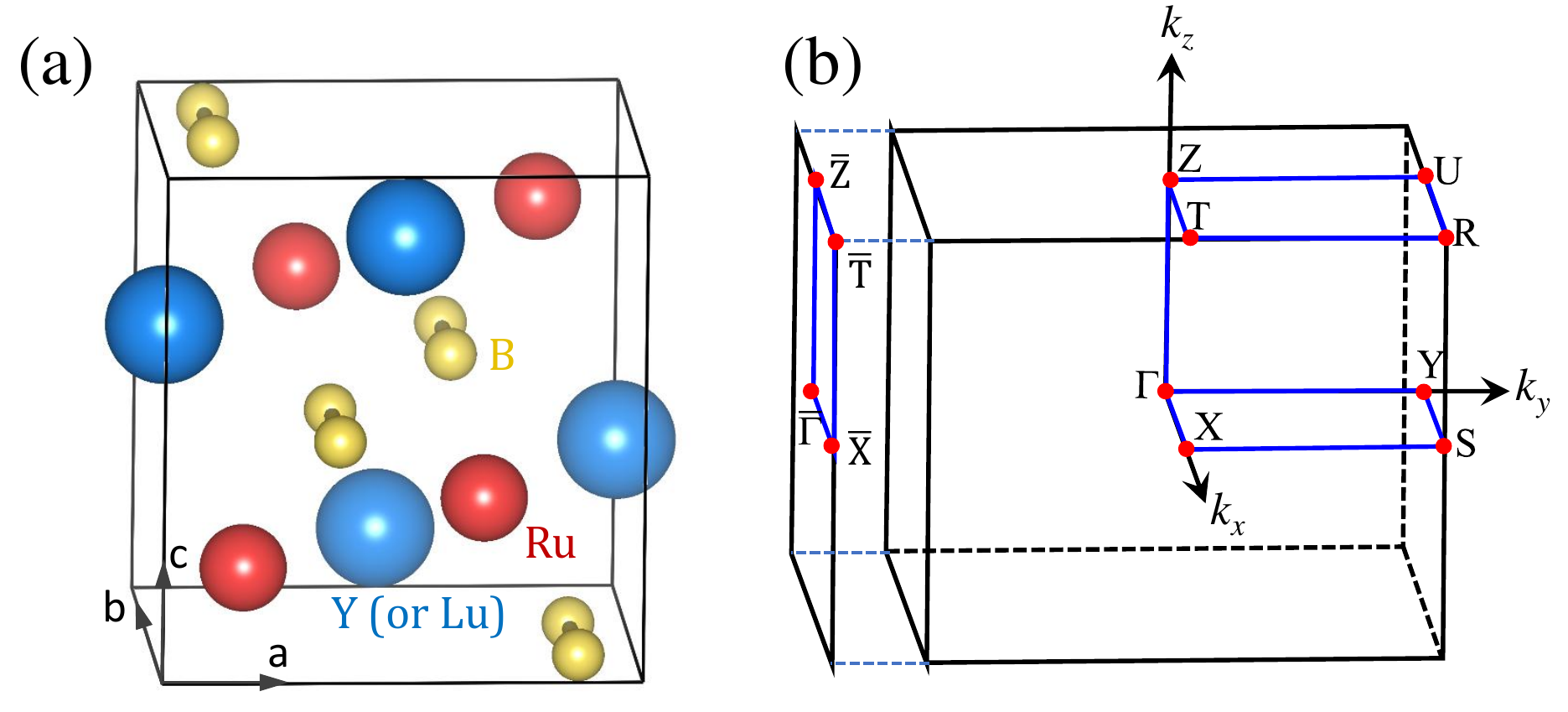}
	\caption{(Color online) (a) Crystal structure of \emph{R}RuB$_{2}$ (\emph{R}=Y, Lu). The blue, red, and golden balls represent \emph{R}, Ru, and B atoms, respectively. (b) Bulk Brillouin zone (BZ) and projected two-dimensional (2D) BZ of the (010) surface. The red dots and blue lines represent the high-symmetry points and paths in the BZ, respectively.}
	\label{fig_structure}
\end{figure}

\section{RESULTS}\label{sec_results}

As shown in Fig.~\ref{fig_structure}(a), the ternary borides \emph{R}RuB$_{2}$ (\emph{R}=Y, Lu) own a nonsymmorphic space group No. 62 (\emph{Pnma}) with the corresponding D$_{2h}$ point group symmetry, which can be generated by the following three symmetry elements: the inversion $\mathcal{P}$, the mirror $\mathcal{M}_{y}$: $(x,y,z)\rightarrow(x,-y+\frac{1}{2},z)$, and glide mirror $\widetilde{\mathcal{M}}_{x}$: $(x,y,z)\rightarrow(-x+\frac{1}{2},y+\frac{1}{2},z+\frac{1}{2})$. Based on the combination of these three symmetry operations, we can obtain another glide mirror: $\widetilde{\mathcal{M}}_{z}$: $(x,y,z)\rightarrow(x+\frac{1}{2},y,-z+\frac{1}{2})$. Here the tilde represents a nonsymmorphic operation, which includes a translation with fractional lattice parameters. Moreover, the \emph{R}RuB$_{2}$ are nonmagnetic, indicating that the time-reversal symmetry $\mathcal{T}$ is preserved~\cite{33BarkerPaul}. Their three-dimensional (3D) bulk Brillouin zone (BZ) with the projected two-dimensional (2D) BZ of the (010) surface are plotted in Fig.~\ref{fig_structure}(b).

\begin{figure*}[!t]
	\centering
	\includegraphics[width=0.7\textwidth]{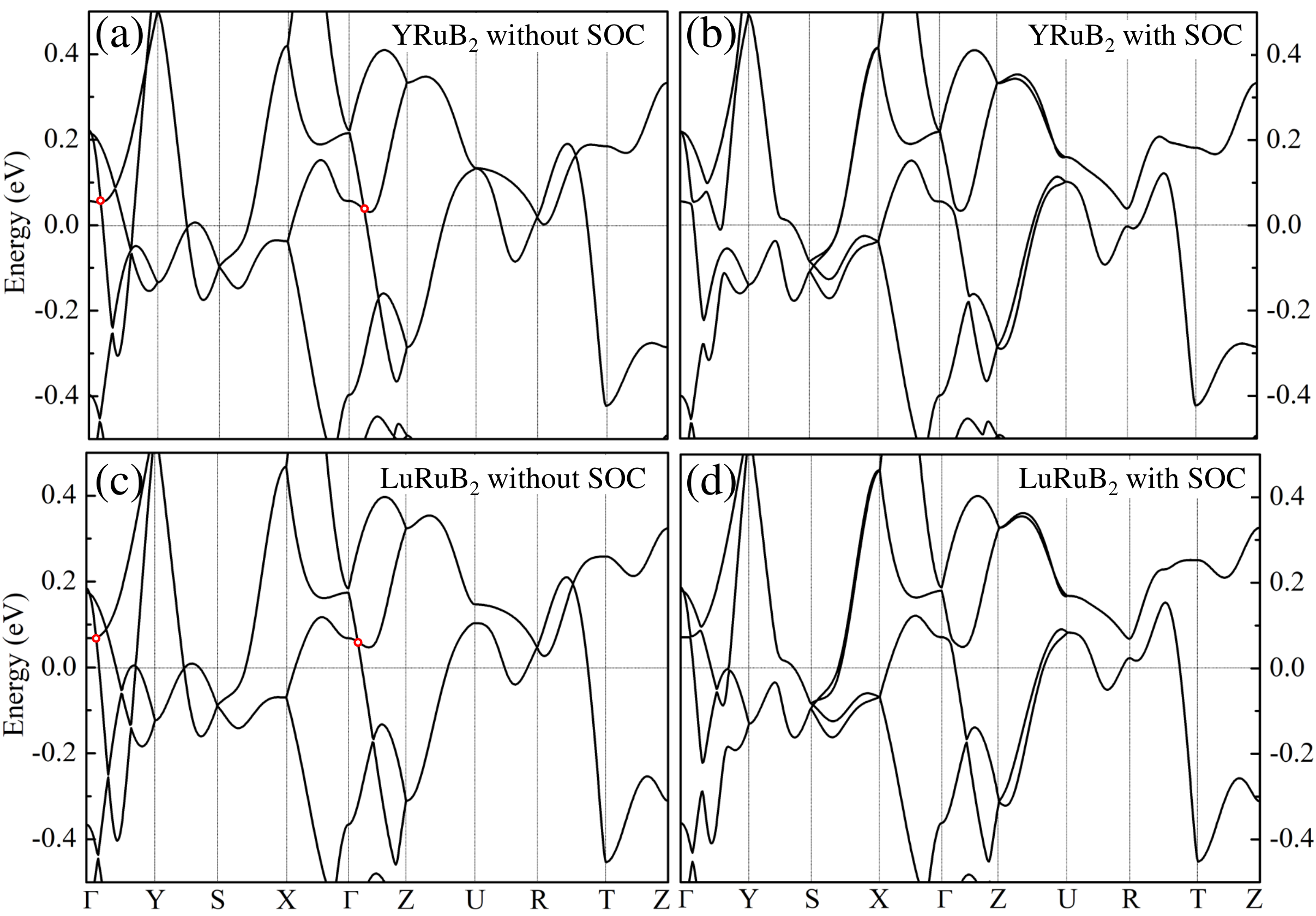}
	\caption{(Color online) Band structures of YRuB$_{2}$ and LuRuB$_{2}$ calculated without and with the spin-orbit coupling (SOC) along the high-symmetry paths in the BZ, respectively.}
	\label{fig_band}
\end{figure*}

The calculated band structures of YRuB$_{2}$ and LuRuB$_{2}$ without the spin-orbit coupling (SOC) are shown in Figs.~\ref{fig_band}(a) and~\ref{fig_band}(c), respectively. One can see that YRuB$_{2}$ and LuRuB$_{2}$ have similar electronic band structures, in which the band crossing points (red circles) along the high-symmetry $\Gamma$-Y and $\Gamma$-Z paths indicate that there is a nodal ring around the $\Gamma$ point in the $k_x=0$ plane protected by the $\widetilde{\mathcal{M}}_{x}$ symmetry (see Supplemental Material (SM) Fig. S1~\cite{34SM}). Since the symmorphic and nonsymorphic space groups have the same irreducible representations in the interior of BZ and the highest rotation symmetry is of two-fold axis for YRuB$_{2}$ and LuRuB$_{2}$, the nodal ring around the $\Gamma$ point in the $k_x=0$ plane will open a complete gap with the inclusion of SOC. Thus both YRuB$_{2}$ and LuRuB$_{2}$ have the band inversion around the $\Gamma$ point [Figs.~\ref{fig_band}(b) and~\ref{fig_band}(d)], which is crucial for the appearance of their topological surface states (TSSs) (We will discuss this point in detail later). We have also examined the correlation effect in the 4\emph{d} electrons of Ru atoms with different Hubbard U values, in which no substantial change is found for the band structure near the Fermi level (see Fig. S2 in the SM~\cite{34SM}).

\begin{figure*}[!t]
	\centering
	\includegraphics[width=0.7\textwidth]{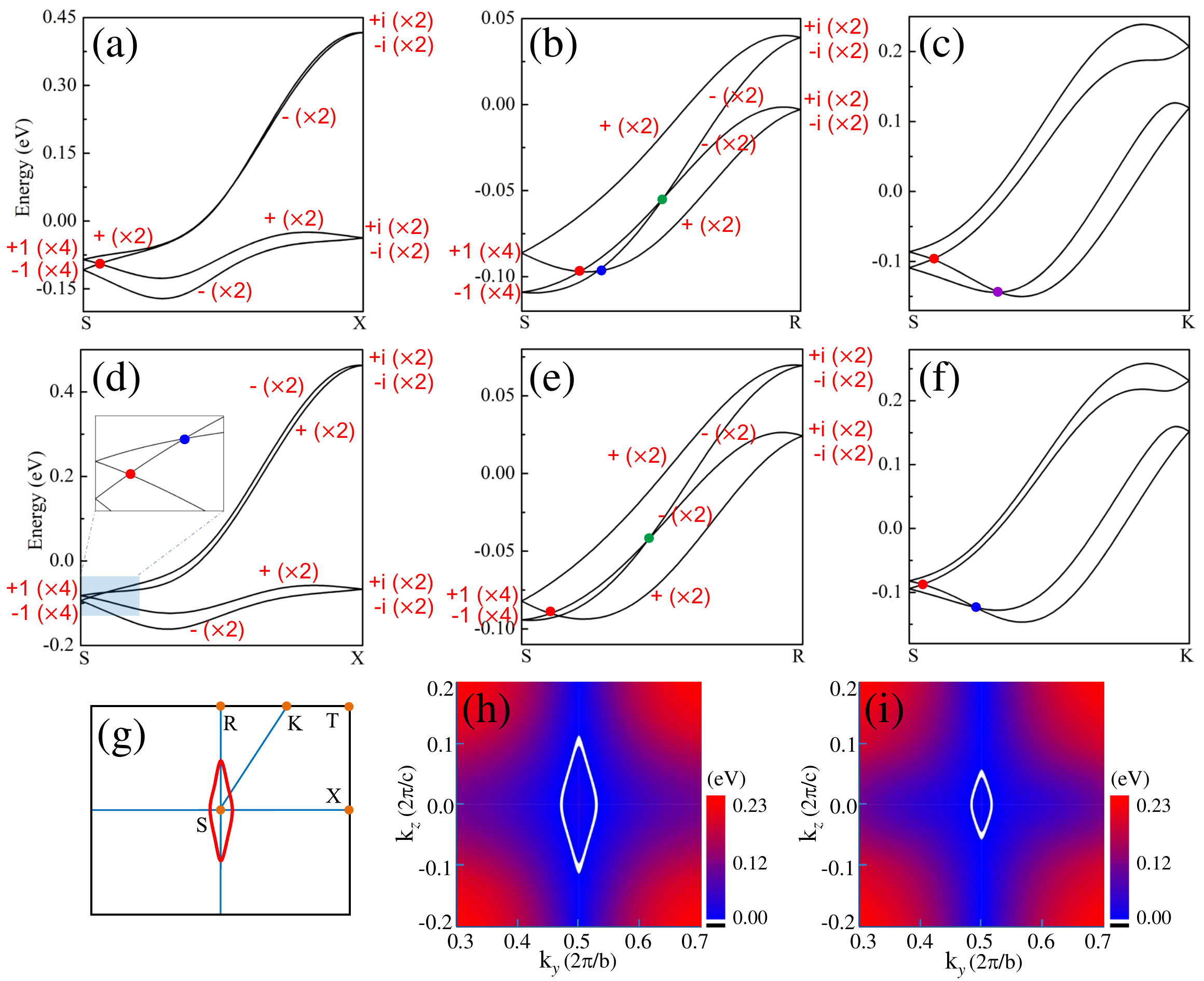}
	\caption{(Color online) The hourglass-type band dispersions of (a-c) YRuB$_{2}$ and (d-f) LuRuB$_{2}$ along paths S-X, S-R, and S-K in the presence of SOC. Here K is an arbitrary point on path T-R, we take the midpoint between T and R for illustration as shown in (c) and (f). The red numbers and signs in the figures represent the eigenvalues $g_{x}$ of $\widetilde{\mathcal{M}}_{x}$, and the numbers in brackets show the degeneracy of the eigenvalues. (g) The schematic figure of the fourfold degenerate Dirac ring formed by the neck points [red dots] of the hybrid hourglass dispersion on the $k_{x}=\pi$ plane. (h-i) The distribution of the hourglass Dirac ring (the white colored rings) surrounding point S for YRuB$_{2}$ and LuRuB$_{2}$, respectively. The color bar shows the local gap between the two crossing bands.}
	\label{fig_largeband}
\end{figure*}

On the other hand, the hourglass-type fermions protected by nonsymmorphic symmetry have been proposed in recent works~\cite{34CavaBernevig,35LiSAYang,36WangHyao}. Interestingly, YRuB$_{2}$ and LuRuB$_{2}$ display an hourglass-type dispersion in a large energy range from -0.17~eV to 0.42~eV along the S-X path as shown in Figs.~\ref{fig_band}(b) and~\ref{fig_band}(d) [the corresponding enlargements shown in Figs.~\ref{fig_largeband}(a) and~\ref{fig_largeband}(d)]. Let's first consider the hourglass-type dispersion along path S-X: ($\pi$,$k_y$,0), where  -$\pi<k_{y}\leq\pi$. This path locates in the $k_x=\pi$ plane which is invariant under $\widetilde{\mathcal{M}}_{x}$, so that each Bloch state $|\Phi_{n}\rangle$ at the path is also an eigenstate of $\widetilde{\mathcal{M}}_{x}$. Since the rotation or mirror operations under the SOC not only act on the spatial coordinates but also on the spin degrees of freedom, we have
\begin{equation}\label{eq_1}
\begin{aligned}
  (\widetilde{\mathcal{M}}_{x})^{2} &= \mathcal{T}_{011}\bar{E} = -e^{-ik_{y}},
\end{aligned}
\end{equation}
in the S-X path, where $\bar{E}$ denotes the 2$\pi$ spin rotation and $\mathcal{T}_{011}: (x,y,z)\rightarrow(x,y+b,z+c)$ is the lattice translation operator. Hence the eigenvalue $g_{x}$ of $\widetilde{\mathcal{M}}_{x}$ must be $\pm$$ie^{-ik_{y}/2}$.  For a state $|\Phi_{n}\rangle$ with an eigenvalue $g_{x}$ of $\widetilde{\mathcal{M}}_{x}$, it has a Kramers partner $\mathcal{PT}|\Phi_{n}\rangle$ satisfying
\begin{equation}\label{eq_2}
\begin{aligned}
  \widetilde{\mathcal{M}}_{x}(\mathcal{PT}|\Phi_{n}\rangle) &= g_{x}(\mathcal{PT}|\Phi_{n}\rangle).
\end{aligned}
\end{equation}
This demonstrates that the Kramers partners $|\Phi_{n}\rangle$ and $\mathcal{PT}|\Phi_{n}\rangle$ in the S-X path share the same eigenvalue $g_{x}$.

At the same time, the S and X points are the time-reversal invariant momenta (TRIM). At the S point: ($\pi$,$\pi$,0), we can get $g_{x}=\pm1$ (we further obtain that the eigenvalue of lower energy is $g_{x}=-1$ and the higher one is $g_{x}=+1$ by computing the traces of the corresponding matrix representations~\cite{37BernevigWang}) and
\begin{equation}\label{eq_3}
\begin{aligned}
  \widetilde{\mathcal{M}}_{x}(\mathcal{T}|\Phi_{n}\rangle) &= g_{x}(\mathcal{T}|\Phi_{n}\rangle),
\end{aligned}
\end{equation}
hence each Kramers partner $|\Phi_{n}\rangle$ and $\mathcal{T}|\Phi_{n}\rangle$ at the S point must have the same eigenvalue $g_{x}$. Meanwhile,
\begin{equation}\label{eq_4}
\begin{aligned}
  \widetilde{\mathcal{M}}_{x}(\mathcal{P}|\Phi_{n}\rangle) &= g_{x}(\mathcal{P}|\Phi_{n}\rangle),
\end{aligned}
\end{equation}
thus we may choose $|\Phi_{n}\rangle$, $\mathcal{T}|\Phi_{n}\rangle$, $\mathcal{P}|\Phi_{n}\rangle$, and $\mathcal{PT}|\Phi_{n}\rangle$ as a degenerate quartet at the S point, which must have the same $g_{x}$.

The similar analysis also applies to the X point. At X: ($\pi$,0,0), since $g_{x}$=$\pm$i, each Kramers partner $|\Phi_{n}\rangle$ and $\mathcal{T}|\Phi_{n}\rangle$ should have opposite eigenvalues $g_{x}$, and the degenerate quartet must be composed of two states with $g_{x}=+i$ and two other states with $g_{x}=-i$. Therefore, there must be a pair switching when going from S to X, which gives rise to an hourglass-type dispersion [see Fig.~\ref{fig_largeband}(a)]. In particular, the neck point (red dot) of hourglass-type dispersion is a fourfold degenerate point for YRuB$_{2}$, which is formed by the enforced band-crossing of two degenerate bands with the opposite eigenvalues $g_{x}$. In contrast, it is not completely an hourglass-type dispersion along the S-X path for LuRuB$_{2}$ [see Fig.~\ref{fig_largeband}(d)], which is caused by the increasing band dispersion due to the strengthened SOC. In addition to the hourglass-type Dirac fermion (red dot), there is a type-II Dirac fermion (blue dot) derived from the two degenerate bands with the opposite eigenvalues $g_{x}$ [shown in the inset of Fig.~\ref{fig_largeband}(d)]. Here the hourglass-type fermion accompanying the type-II Dirac fermion is termed as a hybrid hourglass-type fermion, and the nodal line composed of hybrid hourglass-type fermions as the hybrid hourglass-type nodal line. To our knowledge, the hybrid hourglass-type Dirac fermion found here has not been discussed before.

Likewise, for path S-R: ($\pi$,$\pi$,$k_{z}$), we have
\begin{equation}\label{eq_5}
\begin{aligned}
  (\widetilde{\mathcal{M}}_{x})^{2} &= \mathcal{T}_{011}\bar{E} = e^{-ik_{z}},
\end{aligned}
\end{equation}
on S-R. Thus the eigenvalue $g_{x}$ of $\widetilde{\mathcal{M}}_{x}$ must be $\pm$$e^{-ik_{z}/2}$. For a state $|\Phi_{n}\rangle$, it has a Kramers partner $\mathcal{PT}|\Phi_{n}\rangle$ that satisfies Eq.~\ref{eq_2}. Therefore, the Kramers partners $|\Phi_{n}\rangle$ and $\mathcal{PT}|\Phi_{n}\rangle$ on S-R share the same eigenvalue $g_{x}$. At R: ($\pi$,$\pi$,$\pi$), we can also get $g_{x}$=$\pm$i. Hence at R each Kramers partner $|\Phi_{n}\rangle$ and $\mathcal{T}|\Phi_{n}\rangle$ should have the opposite eigenvalues $g_{x}$, and the degenerate quartet must be composed of two states with $g_{x}=+i$ and two other states with $g_{x}=-i$. Thus, there must be also a pair switching when going from S to R. But, different from the case with the S-X path, the band dispersion of YRuB$_{2}$ and LuRuB$_{2}$ changes dramatically along path S-R [see Figs.~\ref{fig_largeband}(b) and~\ref{fig_largeband}(e)]. This gives rise to multiple band inversions as well as the emergence of the type-I Dirac fermion (red dot) or type-II Dirac fermion (marked by the blue dots). Here we remind that the band-crossing points (green dots) have the same eigenvalue $g_{x}$ of $\widetilde{\mathcal{M}}_{x}$. Thus they will be gapped by a perturbation, except for the one with the protection by a screw axis $\widetilde{\mathcal{C}}_{2z}$. Here we will not discuss it in detail, because our main concern is about the hourglass-type Dirac ring formed by the neck points (red dots) of the hourglass-type dispersion.

The above analysis can be further generalized to any one path connecting S to a point on the boundary lines T-R or X-T in the $k_{x}=\pi$ plane [see Fig.~\ref{fig_largeband}(g)]. The degeneracies on paths T-R or X-T can be explained in a similar way. Here we take an arbitrary K point on path T-R for analysis. At K: ($\pi$,$k_{y}$,$\pi$), we have
\begin{equation}\label{eq_6}
\begin{aligned}
  (\widetilde{\mathcal{M}}_{x})^{2} &= \mathcal{T}_{011}\bar{E} = e^{-ik_{y}},
\end{aligned}
\end{equation}
hence the eigenvalue $g_{x}$ of $\widetilde{\mathcal{M}}_{x}$ must be $\pm$$e^{-ik_{y}/2}$. For a state $|\Phi_{n}\rangle$, it has a Kramers partner $\mathcal{PT}|\Phi_{n}\rangle$ satisfying Eq.~\ref{eq_2}. Thus, the Kramers partners $|\Phi_{n}\rangle$ and $\mathcal{PT}|\Phi_{n}\rangle$ at the K point share the same eigenvalue $g_{x}$. Meanwhile, path T-R keeps invariant with respect to the $\widetilde{\mathcal{M}}_{z}$ operation. The commutation relation between $\widetilde{\mathcal{M}}_{x}$ and $\widetilde{\mathcal{M}}_{z}$ is
\begin{equation}\label{eq_7}
\begin{aligned}
  \widetilde{\mathcal{M}}_{x}\widetilde{\mathcal{M}}_{z} &= -\mathcal{T}_{\bar101}\widetilde{\mathcal{M}}_{z}\widetilde{\mathcal{M}}_{x},
\end{aligned}
\end{equation}
where the negative sign is the result of these operations on spin, i.e., from $\{\sigma_{x},\sigma_{z}\}=0$, so that $\{\widetilde{\mathcal{M}}_{x}, \widetilde{\mathcal{M}}_{z}\}=0$ on path T-R. Then we can get
\begin{equation}\label{eq_8}
\begin{aligned}
  \widetilde{\mathcal{M}}_{x}(\widetilde{\mathcal{M}}_{z}|\Phi_{n}\rangle) &= -g_{x}(\widetilde{\mathcal{M}}_{z}|\Phi_{n}\rangle),
\end{aligned}
\end{equation}
indicating the two states $\widetilde{\mathcal{M}}_{z}|\Phi_{n}\rangle$ and $|\Phi_{n}\rangle$ have the opposite eigenvalues on path T-R. Likewise, at each $k$ point on path T-R, the degenerate quartet of $|\Phi_{n}\rangle$, $\mathcal{PT}|\Phi_{n}\rangle$, $\widetilde{\mathcal{M}}_{z}|\Phi_{n}\rangle$, and $\widetilde{\mathcal{M}}_{z}\mathcal{PT}|\Phi_{n}\rangle$ must be composed of two states ($|\Phi_{n}\rangle$ and $\mathcal{PT}|\Phi_{n}\rangle$) with $+e^{-ik_{y}/2}$ and two other states ($\widetilde{\mathcal{M}}_{z}|\Phi_{n}\rangle$  and $\widetilde{\mathcal{M}}_{z}\mathcal{PT}|\Phi_{n}\rangle$) with $-e^{-ik_{y}/2}$. Therefore, there must be a pair switching when going from S to the arbitrary K point on path T-R (similar for path X-T). Figure~\ref{fig_largeband}(c) shows the hourglass-type band dispersion accompanying an extra Dirac fermion (purple dot) for YRuB$_{2}$, derived from the path connecting S to the point K on T-R. One notes that there is another type-II Dirac fermion (blue dot) along path S-K for LuRuB$_{2}$ in addition to the hourglass-type fermion [see Fig.~\ref{fig_largeband}(f)]. As a result, the neck points (red dots) of the hourglass-type dispersion of YRuB$_{2}$ and LuRuB$_{2}$ are guaranteed and meanwhile there forms a continuous nodal ring encircling point S in the $k_{x}=\pi$ plane [see Fig.~\ref{fig_largeband}(g)]. This is further confirmed by our DFT calculations, as shown in Figs.~\ref{fig_largeband}(h) and~\ref{fig_largeband}(i) for YRuB$_{2}$ and LuRuB$_{2}$ respectively. Thus this hybrid hourglass-type Dirac ring is essentially and entirely determined by the nonsymmorphic space group symmetry. At the same time, this also provides various pair switchings of glide-reflection eigenvalues for the hybrid hourglass-type fermions to form robust Dirac points. In addition, these Dirac points are 80 to 150~meV below the Fermi level, which can be observed by angle-resolved photoemission spectroscopy (ARPES) measurements.

\begin{table}[!b]
\caption{\label{tab:I} Parity products of all occupied band eigenstates at eight TRIM points in the BZs and the Z$_{2}$ indices of YRuB$_{2}$ and LuRuB$_{2}$.}
\begin{center}
\begin{tabular*}{0.95\columnwidth}{@{\extracolsep{\fill}}cccccccccc}
\hline\hline
 &   $\Gamma$  &  R  &  S  &  U  &  T  &  X  &  Y  &  Z  &  ($\nu_0$; $\nu_1$$\nu_2$$\nu_3$) \\
\hline
 YRuB$_{2}$ & - & - & - & - & + & - & - & - & (1; 101) \\
 LuRuB$_{2}$ & - & - & - & - & + & - & - & - & (1; 101) \\
\hline\hline
\end{tabular*}
\end{center}
\end{table}

\begin{figure*}[!t]
	\centering
	\includegraphics[width=0.7\textwidth]{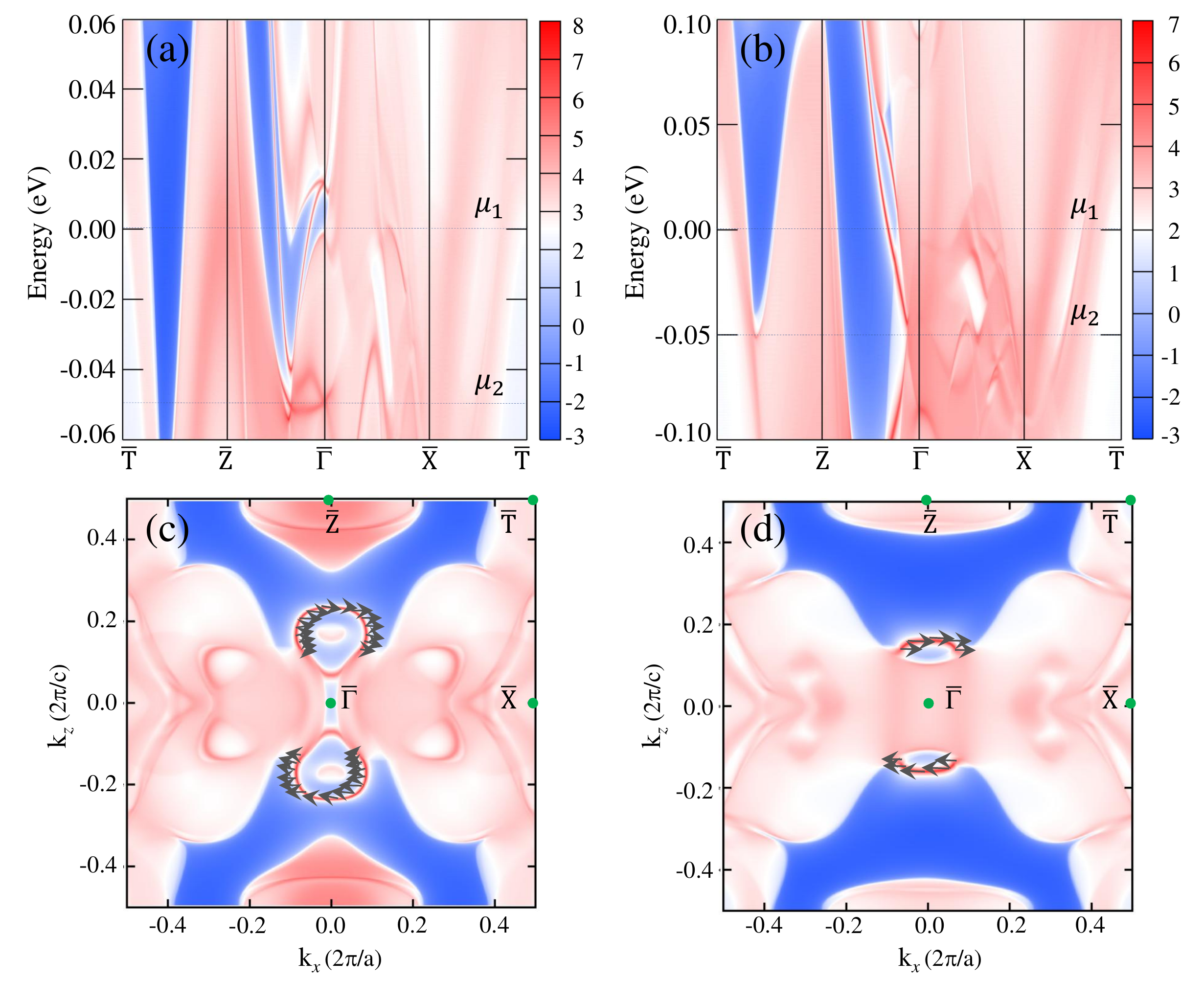}
	\caption{(Color online) Surface energy bands of (a) YRuB$_{2}$ and (b) LuRuB$_{2}$ along the high-symmetry paths in the projected 2D BZ of the (010) surface. The chemical potentials at the Fermi level $\emph{E}_{f}$ ($\mu_{1}$) and 50 meV below the Fermi level ($\mu_{2}$). Surface spectrum of (c) YRuB$_{2}$ and (d) LuRuB$_{2}$ at a fixed energy $\emph{E}_{f}$. Here the black arrows represent the directions of spins.}
	\label{fig_surfaces}
\end{figure*}

As shown above, both YRuB$_{2}$ and LuRuB$_{2}$ have the band inversions around the $\Gamma$ point. Due to the anti-commutation relation of inversion and mirror symmetries, a Bloch state at the TRIM points except for the $\Gamma$ and T points is in quadruple degeneracy with two double-degeneracies marked simultaneously by plus and minus parities. Although \emph{R}RuB$_{2}$ does not have a global band gap in the entire BZ when including the SOC, the topological Z$_{2}$ invariant~\cite{38FuKane} can still be well defined on the $k_y=0$ plane by the product of the parities of all occupied states at the four time-reversal invariant momentum (TRIM) points ($\Gamma$, Z, T, and X) as listed in Table \ref{tab:I}. Our calculation shows that $Z_2=1$, indicating the existence of nontrivial TSSs on the (010) surface in Fig.~\ref{fig_surfaces}. On the other hand, if the nonsymmorphic symmetry above is broken, the hourglass-type Dirac ring will open a band gap, and the band inversion around the $\Gamma$ point will result in a strong topological insulator phase for YRuB$_{2}$ and LuRuB$_{2}$ (see Table \ref{tab:I}). Thus, both YRuB$_{2}$ and LuRuB$_{2}$ have robust TSSs on their surfaces. The calculated TSSs on (010) surfaces of YRuB$_{2}$ and LuRuB$_{2}$ are shown in Figs.~\ref{fig_surfaces}(a) and~\ref{fig_surfaces}(b), respectively. One can note that YRuB$_{2}$ has the Dirac TSSs around the $\bar{\Gamma}$ point on the (010) surface protected by the time-reversal symmetry, while for LuRuB$_{2}$ the Dirac point of surface states at the $\bar{\Gamma}$ point overlaps with the bulk states. Interestingly, both of their TSSs on the (010) surfaces cross the Fermi level and are within the bulk gaps, especially forming single Fermi surfaces [Figs.~\ref{fig_surfaces}(c) and~\ref{fig_surfaces}(d)]. Meanwhile, previous experiments found that both YRuB$_{2}$ and LuRuB$_{2}$ are superconductors~\cite{26Kishimoto}. The superconducting bulks of YRuB$_{2}$ and LuRuB$_{2}$ may induce the equivalent $p$+i$p$ superconductivity in the single TSSs on the (010) surface, which support the Majorana zero modes in the vortex cores. Thus, YRuB$_{2}$ and LuRuB$_{2}$ may be single-compound TSCs satisfying the aforementioned three criteria. Note that if the chemical potential of \emph{R}RuB$_{2}$ moves to 50 meV ($\mu_{2}$) below the Fermi level as shown in Figs.~\ref{fig_surfaces}(a) and~\ref{fig_surfaces}(b), their TSSs will merge into the bulk states (see Fig. S3 in the SM~\cite{34SM}), which could cause the disappearance of the topological superconducting state. This makes the surface topological superconductivity sensitive to the position of the chemical potential~\cite{39NieWang}. However, we also investigate the surface states of YRuB$_{2}$ on the (100) and (001) surfaces, as shown in the SM Fig. S4~\cite{34SM}. We find that the surface states on the (100) and (001) planes are rather complex and these surface states across the Fermi level form double Fermi surfaces, thus we do not discuss them here.

\section{DISCUSSION AND SUMMARY}\label{sec_discussion}

As a new class of topological superconductor candidates with hourglass-type nodal rings, \emph{R}RuB$_{2}$ (\emph{R}=Y, Lu) have many advantages. First, \emph{R}RuB$_{2}$ are such a class of stoichiometric single-compound superconductors (SCs), compared with the doped TIs~\cite{40FuBerg}, the TI/SC heterostructures~\cite{41CLGaoJPXu}, or the pressurized TIs~\cite{42CaoZhang,43Kirshenbaum}, that \emph{R}RuB$_{2}$ can avoid the disorder effect, lattice mismatch, and interface reaction issues. Second, the $\emph{T}_{c}$'s of YRuB$_{2}$ (7.6 K) and LuRuB$_{2}$ (10.2 K) are higher than those of the existing SCs with TSSs, such as Au$_{2}$Pb (1.18 K)~\cite{17Canfield}, PdTe$_{2}$ (1.7 K)~\cite{18NohPark}, PbTaSe$_{2}$ (3.8 K)~\cite{19PJChenHTJeng,20TRChang}, BiPd (3.8 K)~\cite{21Raychaudhuri,22SunWahl}, YD$_{3}$ (0.78~K to 4.72~K)~\cite{23XHTuBTWang}, $\beta$-PdBi$_{2}$ (5.3 K)~\cite{24SakanoIshizaka}. The higher $\emph{T}_{c}$ is helpful for the experimental observation of the MZMs. In addition, the hourglass-type Dirac ring has the drumhead-like surface states~\cite{35LiSAYang}, giving rise to interesting phenomena such as the anomalous Landau level spectrum~\cite{44RMoessner}, the possible surface magnetism/superconductivity~\cite{45LBalents,46Nandkishore}, the unusual optical response~\cite{47EMele}, and the anisotropic electron transport~\cite{48TGlatzhofer}. Importantly, unlike a general nodal line that can be removed by the spin-orbit coupling (SOC), the hourglass-type nodal ring is guaranteed by the nonsymmorphic symmetries and cannot be removed by the SOC.

In summary, based on symmetry analysis and first-principles electronic structure calculations, we predict that \emph{R}RuB$_{2}$ (\emph{R}=Y, Lu) are not only topological superconductors, but also own the hybrid hourglass-type Dirac ring which is protected by the nonsymmorphic space group symmetry. Especially, their topological Dirac surface states remain separated from the bulk states, cross the Fermi level, and preserve single helical spin textures. The superconductivity in bulk YRuB$_{2}$ (7.6 K) and bulk LuRuB$_{2}$ (10.2 K) may further give rise to superconductivity in the topological surface states by the proximity effect. To the best of our knowledge, LuRuB$_{2}$ has the highest $\emph{T}_{c}$ among the stoichiometric single-compound TSC candidates. Thus the \emph{R}RuB$_{2}$ (\emph{R}=Y, Lu) materials may provide a very promising platform for the realization of topological superconductivity and hourglass fermions in future experiments.

\begin{acknowledgments}

We wish to thank Weikang Wu and Xinzheng Li for helpful discussions. This work was supported by the National Key R\&D Program of China (Grants No. 2019YFA0308603 and No. 2017YFA0302903), the National Natural Science Foundation of China (Grants No. 11774422 and No. 11774424), the CAS Interdisciplinary Innovation Team, the Fundamental Research Funds for the Central Universities, and the Research Funds of Renmin University of China (Grants No. 16XNLQ01 and No. 19XNLG13). Y.G. was supported by the Outstanding Innovative Talents Cultivation Funded Programs 2020 of Renmin University of China. P.J.G. was supported by the fellowship of China Postdoctoral Science Foundation (Grant No. 2020TQ0347). Computational resources were provided by the Physical Laboratory of High Performance Computing at Renmin University of China.

\end{acknowledgments}

\end{document}